\documentclass[secnumarabic,aps,prb,reprint]{revtex4-2}
\usepackage[english]{babel}
\usepackage{graphicx}
\usepackage{color}
\usepackage{amsmath}
\usepackage{tabularx}
\usepackage[dvipsnames]{xcolor}

\begin{document}

	\title{Anomalous Zeeman effect in SrTiO$_3$ and its possible all-electric detection}
	\author{Sergei Urazhdin}
	\affiliation{Department of Physics, Emory University, Atlanta, GA, USA.}

\begin{abstract}
We show that the interplay between spin-orbit coupling and cubic symmetry breaking in SrTiO$_3$ results in a highly anomalous Zeeman effect of conduction electrons substantially different among the three conduction sub-bands and strongly dependent on their splitting. This effect can be measured via electrically-driven spin resonance enabled by the interplay between electron hopping and spin-orbit coupling, and enhanced by the near-degeneracy of the conduction sub-bands. The proposed effects can provide a unique insight into the electronic properties of SrTiO$_3$ and its heterostructures.
\end{abstract}

\maketitle

\section{Introduction}\label{sec:intro}

The fascinating electronic and structural properties of strontium titanate, SrTiO$_3$ (STO), have motivated its extensive studies for over half a century. Its heterostructures and interfaces exhibit high-mobility 2D electron gas (2DEG)~\cite{Ohtomo2004}, ferroelectricity~\cite{BURKE1971191}, magnetism~\cite{Coey2016} and unconventional superconductivity (SC)~\cite{Liu2021}. Meanwhile, bulk STO is a paradigmatic quantum paraelectric~\cite{PhysRevB.19.3593} which exhibits SC at record-low electron concentrations~\cite{PhysRev.163.380}. A possible connection between quantum paraelectricity and SC, and more generally the relationship between electronic and lattice degrees of freedom in STO is a long-standing open scientific problem~\cite{PhysRevB.6.4740,Uwe1985,PhysRevB.81.235109,PhysRevB.84.201304,PhysRevB.88.045114,PhysRevB.90.125156,PhysRevB.94.035111,PhysRevB.97.144506,PhysRevMaterials.3.091401,Ahadi2019}. In particular, it remains debated whether SC in STO can be explained by the conventional Bardeen-Cooper-Schrieffer (BCS) phonon-mediated mechanism with electron-phonon coupling potentially enhanced by incipient ferroelectricity~\cite{GASTIASORO2020168107}, or is unconventional as in other complex oxides such as cuprates~\cite{https://doi.org/10.1002/anie.198807351,PhysRevLett.121.127002,PhysRevB.106.224519}.

The unusual electronic properties of STO and their strong coupling to lattice distortions can be traced to the singular structure of its conduction band dominated by three sub-bands derived from the $t_{2g}$ orbitals of Ti. The sub-band splitting and ordering is determined by the interplay between structural symmetry breaking and spin-orbit coupling (SOC)~\cite{PhysRev.135.A1321,PhysRevB.6.4740,Uwe1985,Guo2003, PhysRevB.84.201304,PhysRevB.89.155402,PhysRevMaterials.3.022001}. The electronic properties of each sub-band are strongly affected by the its spin-orbital composition, due to the strong orbitally-selective anisotropy of hybridization between the $t_{2g}$ orbitals via the $p$-orbitals of oxygen atoms~\cite{PhysRev.135.A1321,PhysRevB.6.4718}.

The nature of the lowest-energy states remains debated even for bulk STO. Some studies suggest that at modest electron doping, electronic properties are well-described by a single free electron-like quasi-isotropic band~\cite{PhysRevB.84.205111,PhysRevX.3.021002}, while others indicate a highly anisotropic band structure~\cite{PhysRevB.81.235109,PhysRevB.88.045114}. The complexity of this problem is compounded by the polaronic effects~\cite{Swartz1475} and correlations~\cite{Boovi2020,PhysRevB.81.235109,PhysRevB.106.224519,PhysRevMaterials.7.L030801}, which may render the interpretation in terms of single-particle Bloch states inadequate. Thus, new experimental approaches capable of elucidating the nature of low-energy electronic states in STO are highly desirable.

Here, we propose the possibility to elucidate the properties of low-energy electronic states in STO via Zeeman effect, by measurements of electron spin resonance (ESR)~\cite{YAFET19631}. We show that this effect can provide a unique insight into the interplay between lattice distortions and SOC. To illustrate the proposed approach, we analyze the effects of strain-induced uniaxial anisotropy realized, for example, in the tetragonal phase of bulk STO at $T<T_a$ or in thin films. In the next section, we develop a minimal tight binding model that includes the effects of strain and SOC. In Section~\ref{sec:Zeeman}, we utilize this model to analyze Zeeman effect and show its extreme sensitivity to strain. In Section~\ref{sec:ESR}, we discuss possible approaches to its characterization. We summarize our findings, and the possible implications of the proposed measurements for the electronic properties of STO, in Section~\ref{sec:summary}.

\section{Tight-binding model of conduction band structure}\label{sec:bands}

In this section, we develop a tight-binding model of the lowest-energy conduction band states in STO that includes cubic symmetry breaking and SOC. This model is not intended as a new theoretical solution to the long-standing problem of sub-band ordering and structure. Rather, it provides a simple tractable framework that captures the same essential features of the band structure as other methods~\cite{PhysRev.135.A1321,PhysRevB.6.4740}, while allowing us to analyze Zeeman effect and demonstrate the possibility of its electronic characterization.

The conduction band is dominated by three $t_{2g}$ orbitals of Ti, which can be described by the pseudo-vector $(d_1,d_2,d_3)=(d_{yz},d_{xz},d_{xy})$. In the high-temperature cubic phase and with SOC neglected, the corresponding sub-bands are degenerate at the $\Gamma$-point, but become split at finite wavevectors due to the anisotropy of orbitally-selective hopping. Each orbital $d_m$ hybridizes only in two principal directions complementary to the $m^{th}$ direction, so the corresponding sub-band is non-dispersive in this direction~\cite{Dylla2019}.

The degeneracy of the $t_{2g}$ subbands at the $\Gamma$-point is lifted by cubic symmetry breaking and SOC. At temperatures $T$ above the antifferodistortive phase transition at $T_a=110$~K~\cite{PhysRevLett.21.16}, antiferroelectric distortions reduce the crystal symmetry of bulk STO to tetragonal. Lattice mismatch with the substrate similarly results in tetragonal strain in epitaxial thin films~\cite{PhysRevB.90.125156}. We define the z-axis to be along the tetragonal symmetry direction, and assume that x and y directions remain equivalent. Additional effects of inversion symmetry breaking in heterostructures are discussed below in the context of Rashba effect.

We limit our analysis to tetragonal phase of bulk STO and $(001)$-oriented films uniaxially strained in the direction normal to the substrate, $(111)$-oriented films will be discussed elsewhere. The nearest-neighbor tight-binding hopping Hamiltonian projected on the Ti $t_{2g}$ orbitals is~\cite{Dylla2019,PhysRevB.106.224519}
\begin{equation}\label{eq:H_hop}
		\hat{H}_{hop}=\sum_{\mathbf{n},\mathbf{l},m,s} t_m{(1-\delta_{l,m})\hat{c}^+_{\mathbf{n}+\mathbf{l},m,s}\hat{c}_{\mathbf{n},m,s}},
\end{equation}
where the operator $\hat{c}^+_{\mathbf{n},m,s}$ creates an electron on site $\mathbf{n}$ with spin $s$ in the orbital state $m$,  
$\mathbf{l}$ is a unit vector in one of six principal directions, $t_1=t_2\equiv t<0$, and $t_3=t+\Delta$, with $\Delta$ describing anisotropy due to the uniaxial strain. $\Delta>0$ ($\Delta<0$) describes tensile (compressive) in-plane strain. The next-order hopping contribution mediated by oxygen-oxygen hybridization renormalizes the hopping coefficients but does not compromise orbitally-selective anisotropy of this Hamiltonian for $\Delta\ll t$~\cite{PhysRevB.106.224519}.

The atomic SOC Hamiltonian $\hat{H}=\lambda\hat{\mathbf{L}}\cdot\hat{\mathbf{S}}$ of Ti projected on the $t_{2g}$ subspace can be written as~\cite{PhysRevB.106.224519}
\begin{equation}\label{eq:H_SO}
	\hat{H}_{SO}=-i\frac{\lambda}{2}\sum_{\mathbf{k},m_i,s,s'} e_{m_1m_2m_3}\sigma_{m_1}^{ss'}\hat{c}^+_{\mathbf{k},m_2,s}\hat{c}_{\mathbf{k},m_3,s'},
\end{equation}
where $\hat{\mathbf{S}}$, $\hat{\mathbf{L}}$ are spin and orbital moment operators, respectively, $\lambda\approx18$~meV is the spin-orbit coupling parameter of Ti~\cite{Dunn1961-ce,PhysRevB.88.045114}, $e_{m_1m_2m_3}$ is the Levi-Civita symbol, $\sigma_{m}$ is the $m^{th}$ Pauli matrix, and $\hat{c}^+_{\mathbf{k},m,s}$ is the creation operator of a Block electron with wavevector $\mathbf{k}$ on the orbital $d_m$ with spin projection $s=\pm1/2$ on the z-axis.

The ordering and spin-orbital structure of three $t_{2g}$ sub-bands described by the Hamiltonian
\begin{equation}\label{eq:H}
	\hat{H}=\hat{H}_{hop}+\hat{H}_{SO}
\end{equation}
depends on the sign of $\Delta$ and its magnitude relative to $\lambda$. Both aspects have been debated. For tetragonal distortion in bulk STO, some early calculations estimated $\Delta>0$ substantially larger than $\lambda$~\cite{PhysRevB.6.4740}, while recent estimates show that $\Delta$ is significantly smaller than $\lambda$ and may be negative~\cite{PhysRevB.84.205111,PhysRevB.88.045114}. The scenario $\Delta\ll\lambda$ is consistent with the observed splitting of the lowest-energy subbands in bulk tetragonal STO and its thick films of about $2-5$~meV~\cite{Uwe1985,PhysRevB.88.045114,PhysRevX.3.021002}, which is too small to be explained by SOC. However, a much larger than expected from this interpretation sub-band shift of about $25$~meV across $T_a$ was detected by angular-resolved photoemission~\cite{PhysRevB.81.235109}. To reconcile these contradictory results, it was suggested that single-particle approximations may be insufficient to describe the properties of STO~\cite{PhysRevB.81.235109,PhysRevB.106.224519,PhysRevMaterials.7.L030801}.

For thin STO films strained by heteroepitaxy on lattice-mismatched substrates, both compressive and tensile strain exceeding $1\%$ has been reported~\cite{PhysRevB.90.125156}, which according to calculations for thin STO quantum wells can result in relative sub-band shifts of up to a few tens of meV~\cite{PhysRevMaterials.3.014401}. In 2DEGs at STO interfaces, symmetry-breaking due to the anisotropy associated with hybridization across the interface results in a very large sub-band splitting comparable to $t$~\cite{King2014,omar2021large}. Based on these results, one can expect that different signs of $\Delta$ and relations of its magnitude with $\lambda$ can be realized in different STO heterostructures. 

\begin{figure}
	\centering
	\includegraphics[width=1.0\columnwidth]{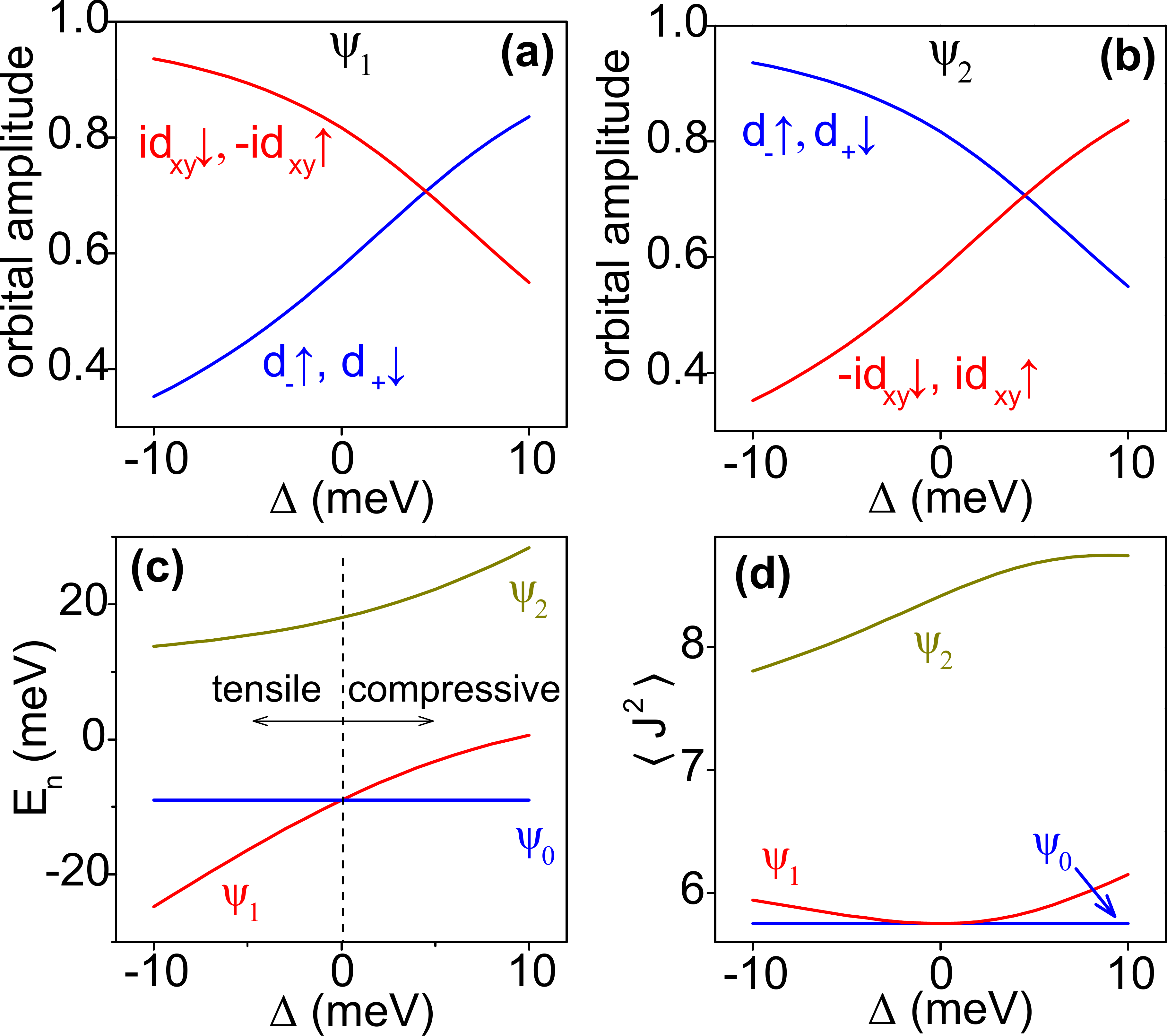}
	\caption{\label{fig:vsDelta} (a),(b) Amplitudes of the two orbital components of $\psi_1$ (b) and $\psi_2$ (c) vs $\Delta$. (c) Energies of the $k=0$ states vs strain parameter $\Delta$. (d) Expectation value of the square of the total atomic angular moment at $k=0$ vs $\Delta$.}
\end{figure}

We use exact diagonalization of the minimal Hamiltonian Eq.~(\ref{eq:H}) at the $\Gamma$-point to analyze the dependence of the spin-orbital structure of the conduction states on $\Delta$, and show that the resulting large variations of Zeeman effect can be used as a sensitive test for the sub-band splitting and ordering. The six $k=0$ Bloch states described by the Hamiltonian Eq.~(\ref{eq:H}) form three Kramers doublets. The first doublet is
\begin{equation}\label{eq:psi0}
	\psi_{0,\uparrow}(k=0)=|d_-\uparrow\rangle, \psi_{0,\downarrow}(k=0)=|d_+\downarrow\rangle,
\end{equation}
where arrows $\uparrow,\downarrow\equiv$ denote the projection of spin on the z-axis, $d_\sigma=-(\sigma d_{xz}+id_{yz})/\sqrt{2}$ are orbital states with projections $\sigma=\pm1$ of orbital moment on this axis.

The remaining two doublets can be written as
\begin{equation}\label{eq:psi1}
	\begin{split}
	\psi_{1,\uparrow}(k=0)=|d_+\uparrow\rangle\cos\xi-i|d_{xy}\downarrow\rangle\sin\xi,\\ 
	\psi_{1,\downarrow}(k=0)=|d_-\downarrow\rangle\cos\xi+i|d_{xy}\uparrow\rangle\sin\xi,
	\end{split}
\end{equation}
and
\begin{equation}\label{eq:psi1}
	\begin{split}
		\psi_{2,\uparrow}(k=0)=|d_+\uparrow\rangle\sin\xi+i|d_{xy}\downarrow\rangle\cos\xi,\\ 
		\psi_{2,\downarrow}(k=0)=|d_-\downarrow\rangle\sin\xi-i|d_{xy}\uparrow\rangle\cos\xi,
	\end{split}
\end{equation}
where trigonometric amplitude parametrization with 
$$
\xi=\tan^{-1}\left[\sqrt{2}(\frac{1}{4}-\frac{\Delta}{\lambda}+\sqrt{\frac{\Delta^2}{\lambda^2}-\frac{\Delta}{2\lambda}+\frac{9}{16}})
\right]$$
is used to simplify normalization. Its value  approaches $\xi=\tan^{-1}\sqrt{2}$ in the limit $\Delta\ll\lambda$. Note that in contrast to $\psi_0$, both $\psi_1$ and $\psi_2$ are spin-orbit entangled, i.e., spin is not a good quantum number. The up/down arrow pseudospin notation labeling the states in a Kramers doublet is retained for convenience. Their orbital composition exhibits a strong dependence on $\Delta$, Figs.~\ref{fig:vsDelta}(a),(b), due to efficient mixing between almost degenerate states. 

The energies $E_n$ of $\psi_n$ relative to the degenerate state in the absence of SOC and strain are shown in Fig.~\ref{fig:vsDelta}(c) for the range $\Delta=-10$ to $10$~meV that based on the calculations for quantum wells reflects the effects of strain ranging roughly from $-1\%$ to $1\%$~\cite{PhysRevMaterials.3.014401}. At $\Delta=0$, $\psi_0$, $\psi_0$, form an orbitally degenerate ground state (g.s.), with $\psi_2$ split off by $3\lambda/2\approx27$~meV. The energy $E_0$ of $\psi_0$ is independent of $\Delta$, while  $E_1-E_0\approx 4\Delta/3$ at $\Delta\ll\lambda$, consistent with \textit{ab initio} calculations for bulk STO~\cite{PhysRevB.84.205111}. Thus, $\psi_1$ is the g.s. at $\Delta<0$, and $\psi_0$ is the g.s. at $\Delta>0$. Since spin is a good quantum number for $\psi_0$ but not $\psi_1$, depending on the sign of $\Delta$, the lowest-energy sub-band may (if it is $\psi_0$) or may not (if it is $\psi_1$) support spin transport, with strain acting as an abrupt switch. This property can be used to distinguish between the two possible lowest sub-band orderings.

None of these states are eigenstates of the total atomic angular moment $\hat{\mathbf{J}}=\hat{\mathbf{L}}+\hat{\mathbf{S}}$ on the full space of five d-orbitals. As a consequence, they are not characterized by half-integer moments. The composition and thus the moment of $\psi_0$ is independent of $\Delta$, with $J^2=5\frac{3}{4}$, Fig.~\ref{fig:vsDelta}(d). At $\Delta=0$, the state $\psi_1$ is characterized by the same moment, consistent with their degeneracy as the eigenstates of $\hat{\mathbf{L}}\cdot\hat{\mathbf{S}}$, while the moment of $\psi_3$ is significantly larger. While the spin-orbital composition of $\psi_1$ is strongly dependent on $\Delta$ its moment only slightly increases at finite $\Delta$. Naively, one would expect a similarly weak variation of response to magnetic field, which is shown below to not be the case. This surprising result is a consequence of complete suppression of Zeeman effect at $\Delta=0$, resulting in large relative variations with $\Delta$.

The electrically-driven ESR discussed below relies on the wavevector dependence of spin-orbital composition of the conduction states. At small $\mathbf{k}=(k_x,k_y,k_z)$, the effects of hopping can be treated as a perturbation of $\psi_{0}(k=0)$. We focus on the lowest two sub-bands expected to dominate electronic properties up to doping $n\approx2\times10^{19}$~cm$^{-3}$, with only one sub-band populated at doping of up to almost $10^{18}$~cm$^{-3}$ in bulk STO~\cite{PhysRevX.3.021002,PhysRevLett.112.207002,PhysRevMaterials.3.022001}. Larger sub-band splitting in strained thin films can shift these critical doping levels up.

To the lowest nontrivial order in $k$, the spin-orbital composition of the sub-band derived from $\psi_0$ is
\begin{equation}\label{eq:psi0_k}
\begin{split}
	\psi_{0,\uparrow}(\mathbf{k})=|d_-\uparrow\rangle+f_{10}(\mathbf{k})|d_+\uparrow\rangle+if_{20}(\mathbf{k})|d_{xy}\downarrow\rangle,\\
	\psi_{0,\downarrow}(\mathbf{k})=|d_+\downarrow\rangle+f_{10}(\mathbf{k})|d_-\downarrow\rangle-if_{20}(\mathbf{k})|d_{xy}\uparrow\rangle,
\end{split}
\end{equation}
where 
\begin{equation}\label{eq:f0}
	\begin{split}
		f_{10}=\frac{a^2t(k_x^2-k^2_y)}{2}\left[\frac{\cos^2\xi}{E_1-E_0}+\frac{\sin^2\xi}{E_2-E_0}\right],\\
		f_{20}=\frac{a^2t(k_x^2-k^2_y)\sin2\xi}{4}\left[\frac{1}{E_2-E_0}-\frac{1}{E_1-E_0}\right].
	\end{split}
\end{equation}

The spin-orbital composition of the sub-band derived from $\psi_1$ is
\begin{equation}\label{eq:psi1_k}
	\begin{split}
		\psi_{1,\uparrow}(\mathbf{k})=\psi_{1,\uparrow}(0)+f_{11}\psi_{2,\uparrow}(0)+f_{21}(\mathbf{k})|d_-\uparrow\rangle,\\
		\psi_{1,\downarrow}(\mathbf{k})=\psi_{1,\downarrow}(0)+f_{11}\psi_{2,\downarrow}+f_{21}(\mathbf{k})|d_+\downarrow\rangle,
	\end{split}
\end{equation}
where 
\begin{equation}\label{eq:f1}
	\begin{split}
		f_{11}=-\frac{a^2tk_z^2\cos\xi\sin\xi}{E_2-E_1}\\
		f_{21}=\frac{a^2t(k_x^2-k^2_y)\cos\xi}{2(E_0-E_1)}.
	\end{split}
\end{equation}
These expressions show that hopping results in mixing among states with different orbital compositions and spins. Mixing between the two lowest sub-bands is scaled by a large parameter $t/(E_1-E_0)$, resulting in a strong wavevector dependence of spin-orbit composition.

The subband derived from $\psi_0$ is characterized by the effective masses $m_x=m_y=2m^*$, $m_z=m^*$, where $m^*=-\hbar^2/2ta^2$, while for the subband derived from $\psi_1$ they are $m_x=m_y=(\sin\xi+0.5\cos\chi)m^*$, $m_z=m^*$. Using the in-plane effective mass $1.8m_e$ estimated from the Nernst effect measurements of bulk STO~\cite{PhysRevX.3.021002} and assuming that $\psi_0$ is the lowest sub-band in this case, we obtain $t\approx 260$~meV. These estimates do not include polaronic effects that can result in mass renormalization~\cite{Swartz1475}.

Since the splitting $(E_0-E_1)/t$ of the two lowest sub-bands relative to the hopping amplitude is small, as is generally expected for bulk STO, a strong wavevector dependence of spin-orbit composition of the sub-bands is expected. For instance, using $\Delta=3$~meV, we estimate $f_{20}\sim0.01$ at $k_y\approx0.01a^{-1}$, and a substantial amplitude $f_{20}\sim0.1$ is reached at $k_y\approx0.03a^{-1}$. Strong dependence of spin-orbital composition on the wavevector results in large variations of electronic properties throughout the Fermi sea even at light doping. For instance, at $n=10^{17}$~cm$^{-3}$, we estimate $|f_{ij}|\sim 0.1$ at the Fermi surface for $\Delta=3$~meV.

\begin{figure}
	\centering
	\includegraphics[width=\columnwidth]{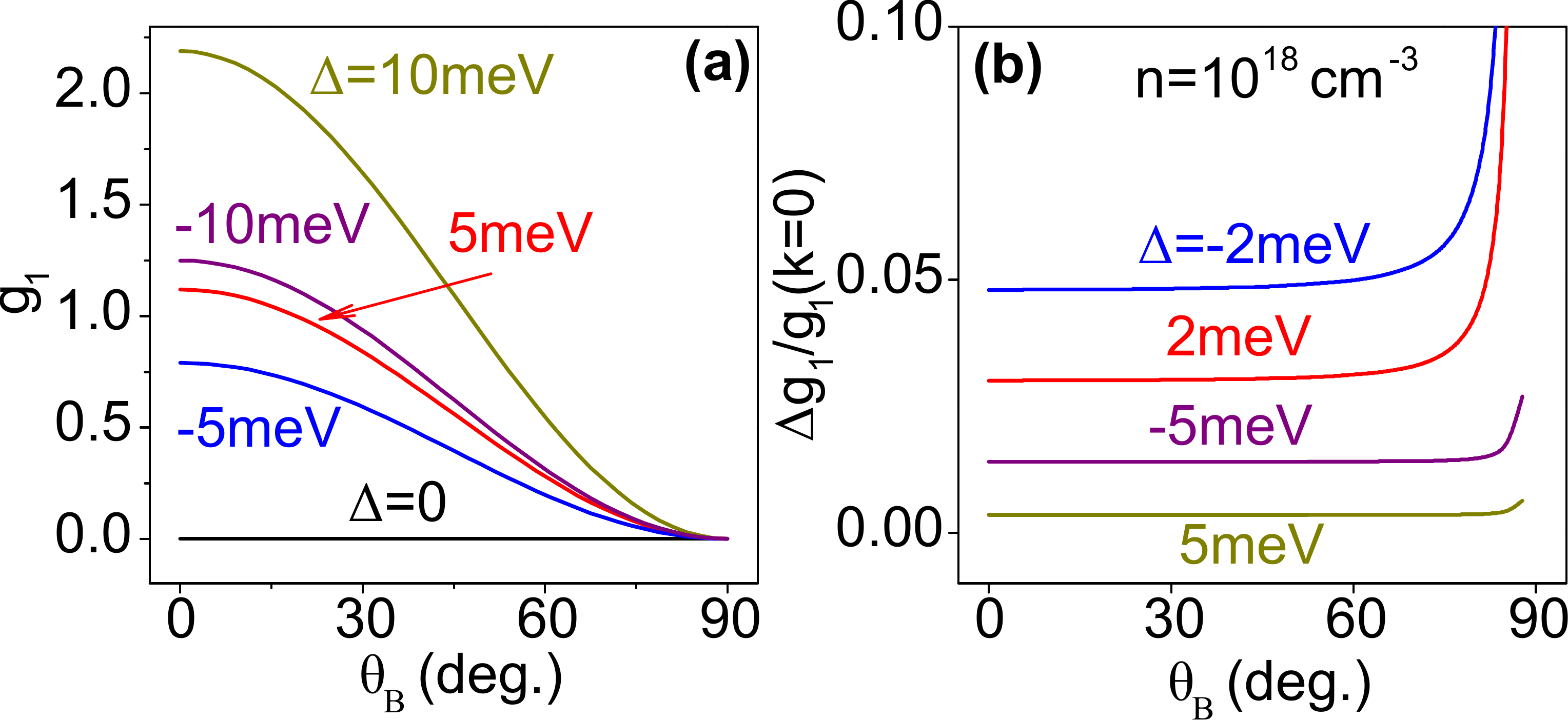}
	\caption{\label{fig:g-factor} (a) g-factor vs polar field angle for the states $\psi_1(k=0)$, at the labeled values of $\Delta$. (b) Relative broadening of the g-factor for the subband $\psi_1(k)$ vs polar field angle, at the labeled values of $\Delta$, for doping $n=10^{18}$~cm$^{-3}$.}
\end{figure}

\section{Zeeman effect and g-factors}\label{sec:Zeeman}


Anomalous Zeeman effect can be generally expected for the conduction band states discussed above due to their non-trivial spin-orbital composition. The Zeeman Hamiltonian is $\hat{H_Z}=\mu_B\mathbf{B}\cdot(2\hat{\mathbf{S}}+\hat{\mathbf{L}})$, where $\mathbf{B}$ is magnetic field and $\mu_B$ is the Bohr magneton. Field-induced splitting $\Delta E_B$ between the sub-band states can be characterized by the spectroscopic g-factor $g=\Delta E_B/\mu_BB$, which generally depends on the field direction.

At $k=0$, the states $\psi_{0}$ Eq.~(\ref{eq:psi0}) are characterized by a vanishing g-factor $g_0=0$, regardless of strain. A very small but finite value was predicted by the \textit{ab initio} calculations~\cite{PhysRevB.84.205111}, consistent with the SOC-induced admixture of the $d_{3z^2-1}$ orbital of the order $\lambda/\Delta E_{eg-t2g}\sim10^{-2}$, where $E_{eg-t2g}\sim2$~eV is crystal-field splitting between $e_g$ and $t_{2g}$ orbitals.

The states $\psi_{1}$ are characterized by the g-factor
\begin{equation}\label{eq:g_1_Delta}
g_1=2|cos\theta(3\cos^2\chi-1)|,
\end{equation}
where $\theta$ is the polar angle of the magnetic field $\mathbf{B}$. At $\Delta\ll\lambda$ $g_1$, $\cos\xi=1/\sqrt{3}$, and the g-factor vanishes. 
At finite $\Delta$, it still vanishes for in-plane field, but reaches significant values for out-of-plane field even at small $\Delta$, Fig.~\ref{fig:g-factor}(a). For instance, it exceeds $2$ for the normal field direction at $\Delta=10$~meV.

This extreme anisotropy of the g-factor and its strong dependence on strain are contrasted with strain-independent isotropically vanishing g-factor of the states $\psi_0$. These properties suggest a simple approach to the characterization of sub-band structure in STO. Such tests will require an experimental verification of the relationship among strain, sub-band splitting, and the Zeeman effect. This can be accomplished by controllable variations of strain by stress in free-standing films~\cite{Lu2016}, or electric bias-controlled strain in films grown on lattice-matched piezoelectric substrates such as Ba$_x$Sr$_{1-x}$TiO$_3$~\cite{Garten2016}.

We now consider the dependence of Zeeman effect on the wavevector, which results from the variation of the spin-orbtial composition described for the two lowest subbands by Eqs.~(\ref{eq:psi0_k}), (\ref{eq:psi1_k}). Zeeman Hamiltonian on the states $\psi_{1,s}(\mathbf{k})$ is
\begin{equation}\label{eq:H_Zeeman_psi1}
\begin{split}
	\hat{H}_{Z,1}=2\mu_BB_z\sigma_z[3\cos^2\chi-1-f_{11}(3\sin^2\xi-1)]\\
	+\mu_B(B_x\sigma_x+B_y\sigma_y)(\cos\chi-\sqrt{2}\sin\xi) f_{21}.
\end{split}
\end{equation}
The first finite-wavevector contribution with amplitude $f_{11}$ describes a modulation of the g-factor without changing its anisotropy. On the other hand, the contribution with amplitude $f_{21}$ results in a finite Zeeman splitting by in-plane field, providing a mechanism for the electronically-driven ESR discussed in the next section. Both contributions result in an inhomogeneous broadening of the g-factor due to the finite Fermi wavevector and/or thermal broadening of electron wavevector distribution. We define g-factor broadening as $\Delta g=\langle |g(\mathbf{k})-g(0)|\rangle$, where averaging is performed over the Fermi sea. Figure~\ref{fig:g-factor}(b) shows the dependence of relative broadening $\Delta g/g(0)$ on the polar field angle estimated for several values of $\Delta$ at $n=10^{18}$~cm$^{-3}$, obtained using a perturbative expansion of $g_{1}(\mathbf{k})$ with respect to $f_{11}$, $f_{21}$ under the assumption that $\psi_1$ is the lowest sub-band and the population of the other sub-bands is negligible. While this expansion becomes accurate only at $\Delta g/g(0)\ll1$, it provides a semi-qualitative picture for the expected broadening effects even if this condition is not satisfied. For doping $n=10^{18}$~cm$^{-3}$ shown in Fig.~\ref{fig:g-factor}(b), broadening is only a few percent at realistic values of $\Delta$, except for the divergence as the field approaches in-plane direction caused by vanishing $g_1(k=0)$. It scales with doping as $n^{5/3}$ for field direction not too far from the film normal, and approaches unity at $n\gtrsim10^{19}$~$cm^{-3}$, constraining $g$-factor characterization to light doping.

Thermal broadening of the Fermi surface leads to a similar inhomogeneous broadening effect. Thermal broadening scales with temperature $T$ as $\Delta g\propto T^{5/2}$. We estimate that at $T=100$~K it is similar to the doping-induced broadening at $n=2\times10^{18}$~cm$^{-3}$. At room temperature $T=300$~K, thermal broadening effects are equivalent to doping $n=3\times10^{19}$~cm$^{-3}$.

\begin{figure}
	\centering
	\includegraphics[width=0.5\columnwidth]{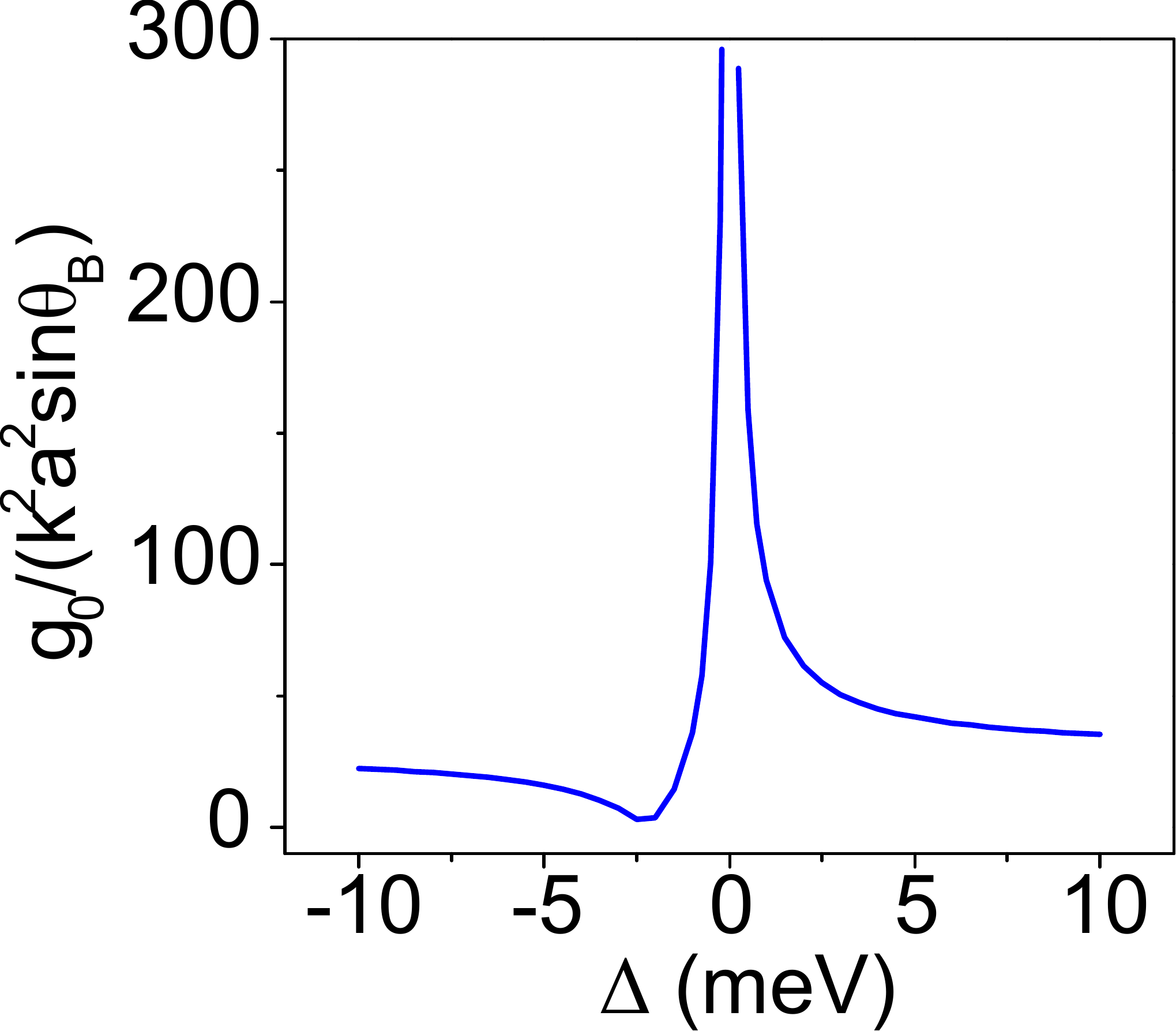}
	\caption{\label{fig:g_0} g-factor vs $\Delta$ for $\psi_0(\mathbf{k})$, calculated using Eq.~(\ref{eq:H_Zeeman_psi0}) and normalized by $(ka)^2\sin\theta_B$, with  $\mathbf{k}$ in the x- or y-direction.}
\end{figure}

To the lowest order in $k$ Zeeman Hamiltonian for the sub-band $\psi_0$ is
\begin{equation}\label{eq:H_Zeeman_psi0}
	\hat{H}_{Z,0}=2\mu_B(B_x\sigma_x+B_y\sigma_y)(f_{10}+f_{20}/\sqrt{2}).
\end{equation}
In the considered approximation, Zeeman splitting vanishes for normal field, but at finite wavevectors becomes finite for in-plane field. This anisotropy of Zeeman splitting is opposite to that identified above for $\psi_1$, providing a simple test for the sub-band ordering. The values of the g-factor calculated using the Zeeman splitting  Eq.~(\ref{eq:H_Zeeman_psi0}) depend non-monotonically on $\Delta$ due to the sub-band splitting-dependent interplay between four spin-orbital pairs that contribute to this effect, Fig.~\ref{fig:g_0}. It exhibits a peak at small positive $\Delta$ close to that believed to describe bulk STO. Using this peak value, we estimate that $g\sim0.7$ is reached at the Fermi surface along the x- and y-directions at $n=10^{18}$~cm$^{-3}$. Since the g-factor vanishes at $k=0$, ESR for this sub-band should be manifested by a very broad peak with the characteristic frequencies that scale as $n^{2/3}$, shifting as $\sin\theta_B$ as the field is rotated in the vertical plane.

\section{Approaches to Zeeman effect characterization}\label{sec:ESR}

In this section, we outline the possible approaches to the characterization of anisotropic g-factors predicted in the previous section. Microwave-absorption ESR, the standard method for g-factor characterization~\cite{YAFET19631}, may be insufficiently sensitive for lightly doped bulk STO or thin films. Indeed, to the best of our knowledge, only impurity ESR at heavy doping has been detected by this technique~\cite{Glinchuk2001}. Zeeman splitting can be characterized via Nernst effect, as was demonstrated for STO in Ref.~\cite{PhysRevX.3.021002}. It may be also possible to extract the g-factor from Shubnikov-de Haas~\cite{JayGerin1979,PhysRevB.88.045114}
or de Haas-van Alphen effects~\cite{PhysRevB.65.184426}.

We now consider an alternative possibility of spin resonance characterization by using the electrical dipole spin resonance (EDSR) effect induced by SOC~\cite{RASHBA1991}. The small values of the expected g-factors may allow all-electrical EDSR driving and detection using the standard electronic microwave techniques, which may be particularly suitable for thin films.

\textit{EDSR in inversion-symmetric systems.} We outline a driving mechanism conceptually analogous to the non-parabolicity mechanism in conventional EDSR of inversion-symmetric systems~\cite{RASHBA1991}. Conventional EDSR relies on the wavevector dependence of orbital mixing between conduction and valence bands, which is inversely proportional to the bandgap and is efficient in heavy-element narrow-gap semiconductors such as InSb where SOC is comparable to the bandgap~\cite{PhysRevLett.51.134}. STO does not meet these criteria. Moreover, the spin-orbital composition of the sub-bands is itself determined by SOC, with hopping at small $k$ playing the role of perturbation. This relationship is opposite to that in the conventional EDSR. However, the same fundamental driving mechanism associated with the dependence of Zeeman Hamiltonian on the wavevector is expected for STO. Oscillating electric field modulates electron wavevector, which acts as an effective  ac magnetic field driving ESR. We argue that EDSR should be highly efficient in STO due to the strong wavevector dependence of spin-orbital composition of the conduction band states and high electron mobility.

\begin{figure}
	\centering
	\includegraphics[width=0.8\columnwidth]{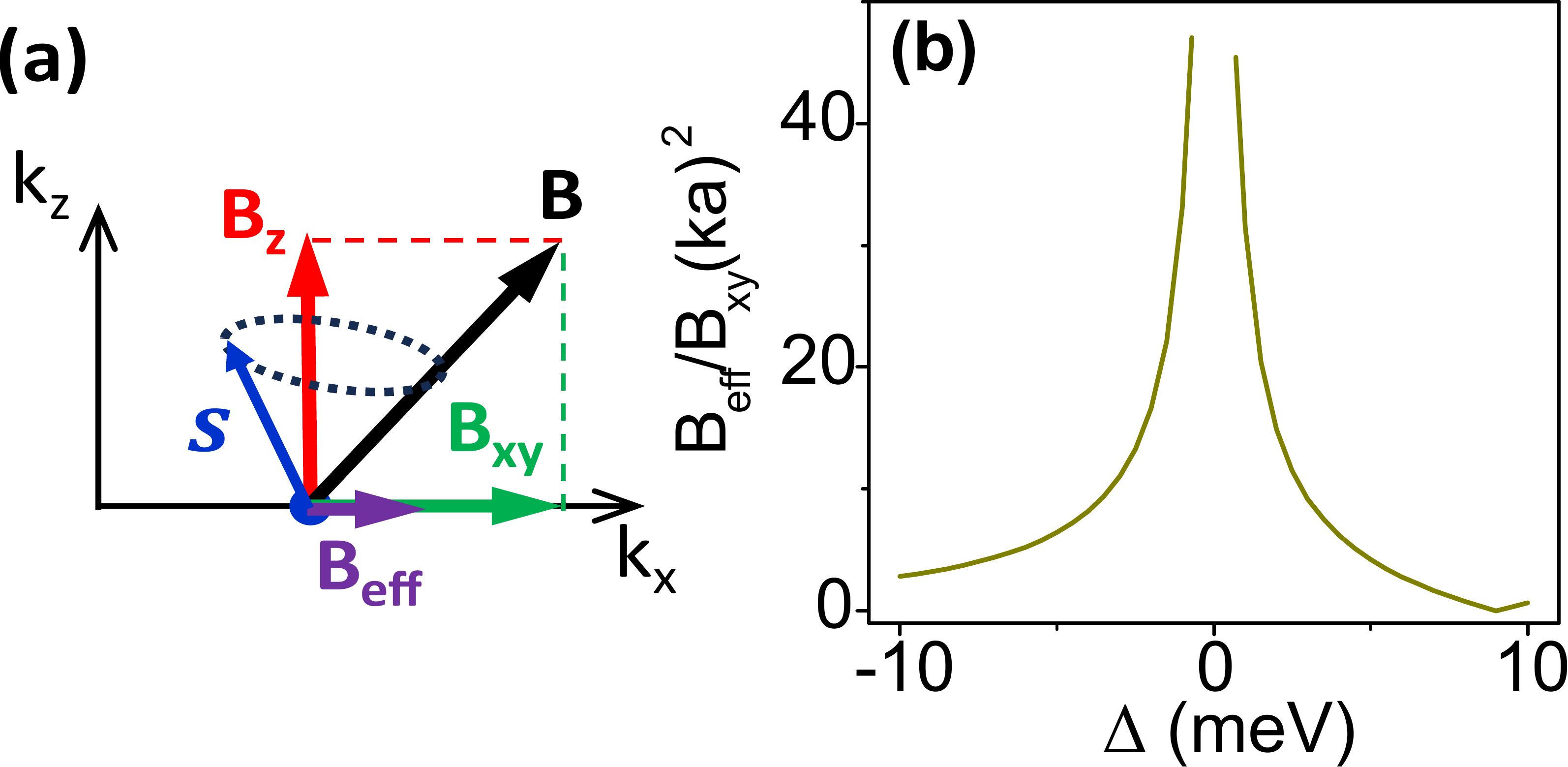}
	\caption{\label{fig:ESR_psi1} (a) Mechanism of EDSR for the sub-band $\psi_1$, (b) Dependence of normalized effective driving field on $\Delta$.}
\end{figure}

The possibility to produce an effective transverse ac magnetic field by modulating the Bloch state wavevector for the sub-band $\psi_1(\mathbf{k})$ is apparent from the form of the Zeeman Hamiltonian Eq.~(\ref{eq:H_Zeeman_psi1}). Assume that dc magnetic field is tilted with respect to the field normal, with the in-plane component $B_{xy}=B\sin\theta_B$. At $n\lesssim10^{18}$~cm$^{-3}$ and $T<\Delta/k_B$, a single sub-band is populated~\cite{PhysRevLett.112.207002}. At light doping and cryogenic temperatures, in the lowest order approximation the Fermi wavevector can be neglected, then only the z-component of the field, the first term in Eq.~(\ref{eq:H_Zeeman_psi1}), contributes to Zeeman effect. Neglecting cyclotron motion effects in the diffusive regime, an ac electric field $E\cos\omega_0 t$ along the x-axis modulates the x-component of the wavevector as $k_x=-m_x\mu E/\hbar$, where $\mu$ is the electron mobility. According to Eq.~(\ref{eq:f1}), this results in the oscillation of the function $f_{21}$ at frequency $2\omega$, producing an effective ac in-plane field $B_{eff}=(\cos\chi-\sqrt{2}\sin\xi)f_{21}B_{xy}$ in the transverse ESR geometry, Fig.~\ref{fig:ESR_psi1}(a).

A pulsed ac electric field at half the frequency corresponding to the Zeeman splitting, $\omega=2\mu_BB_z(3\cos^2\chi-1)/\hbar$, can be utilized to drive transient Rabi spin oscillations. Alternatively, cw field can be used to produce a steady-state spin precession with amplitude determined by the spin relaxation rate. At cryogenic temperatures, lightly doped STO exhibits a high electron mobility of up to $\mu=10^4$~cm$^2$/V$\cdot$s~\cite{doi:10.1146/annurev-conmatphys-031218-013144}, enabling large modulations of wavevector by electric field. Additionally, the function $f_{21}$ is scaled by a large parameter $t/(E_0-E_1)$ [See Eq.~(\ref{eq:f1})], making this excitation mechanism particularly efficient in STO. Figure~\ref{fig:ESR_psi1}(b) shows the calculated dependence of the effective ac magnetic field  on $\Delta$, normalized by the applied in-plane field and the amplitude of wavevector oscillation. As expected from the above arguments, the effective driving field is maximized at small $\Delta$, due to the increased efficiency of mixing between two lowest-energy subbands.

A similar mechanism is expected for the sub-band $\psi_0$. In the considered approximation Zeeman splitting vanishes at the $\Gamma$-point for this sub-band. A small but finite g-factor is expected for the contributions neglected in this approximation such as  $e_g$-$t_{2g}$ orbital mixing by SOC resulting in an admixture of the $d_{3z^1-1}$ orbital, and from inversion symmetry breaking. Similarly to $\psi_1$, modulation of the wavevector by in-plane electric field results in an effective ac magnetic field proportional to the in-plane component of dc magnetic field, as described by the Zeeman Hamiltonian Eq.~(\ref{eq:H_Zeeman_psi0}).
Detailed predictions for this case require analysis of the higher-order contributions to $\psi_0$ neglected in our model.


\textit{Electrical detection of EDSR.} Spin resonance provides a mechanism for electromagnetic energy relaxation, which is detected as a peak in the ac field  absorption in both standard ESR~\cite{YAFET19631} and in EDSR~\cite{RASHBA1991}. The same general mechanism is expected to result in an increase of the imaginary component of electrical impedance in electrically-driven resonance discussed above.

We propose an alternative route based on the Hall voltage due to the oscillating anomalous Hall effect (AHE) produced by the dynamical spin polarization. AHE is associated with spin-dependent anomalous velocity of the Bloch states (intrinsic contribution) or chirality of scattering on impurities (extrinsic contribution)~\cite{RevModPhys.82.1539}. Such effects can be particularly significant in materials where SOC dominates the structure of nearly degenerate conduction band, as in the considered system. Analysis of chiral transport is beyond the scope of this work. Nevertheless, observation of extraordinary thermal Hall effect in STO interpreted in terms of chiral spin-phonon scattering suggests the possibility of a reciprocal effect, chiral spin transport due to scattering on phonons, i.e. a large extrinsic AHE may be expected in the spin-polarized state~\cite{PhysRevLett.124.105901}. The possibility of a large intrinsic AHE of spin-polarized electrons is supported by the prediction~\cite{PhysRevMaterials.3.014401} and observation~\cite{Trier2019} of a large spin Hall effect in STO heterostructures, since AHE and SHE originate from the same chiral transport mechanism. Regardless of the possible mechanism of AHE in spin-polarized STO, the anomalous Hall current density can be written as~\cite{RevModPhys.82.1539}
\begin{equation}\label{eq:AHE}
	\mathbf{j}_{AH}=2\sigma_{AH}\mathbf{E}\times\mathbf{s},
\end{equation}
where $\sigma_{AH}$ is the AHE conductivity, $\mathbf{E}$ is electric field, and $\mathbf{s}$ is the expectation value of the (pseudo)spin vector. According to this equation, mixing between driving ac electric field at frequency $\omega$ and pseudospin oscillating at frequency $2\omega$ is expected to produce AHE current at the frequencies $\omega$ and $3\omega$. In the limit of negligible in-plane Zeeman effect, steady-state precession around the z-axis results in an oscillating AHE in the z-direction, presenting a challenge for its detection in thin films. On the other hand, finite-momentum contribution to $\psi_1$ produces a non-vanishing in-plane contribution to Zeeman effect, as described by the last term in Eq.~(\ref{eq:H_Zeeman_psi1}), which for a tilted field results in the oscillation of the z-component of pseudospin. The latter can be detected as an in-plane AHE signal, which is amenable to detection in thin films.

\textit{Electric dipole resonance via Rashba mechanism}. The most efficient EDSR has been observed in systems with broken inversion symmetry, due to the momentum-dependent effective spin-orbit field described by the Rashba Hamiltonian~\cite{RASHBA1991}
\begin{equation}\label{eq:RE}
	\hat{H}_{R}=\alpha_R\hat{z}\cdot[\mathbf{k}\times\mathbf{\sigma}],
\end{equation}
where $\mathbf{\sigma}$ is the vector of Pauli matrices in spin space~\cite{Rashba2005}. In STO films, inversion symmetry is usually broken by the effective electric field at interfaces, and/or by the static ferroelectric distortions stabilized by the interfaces and strain in thin films~\cite{Uwe1976,SalmaniRezaie2020}.

Rashba effect results from the effective electric field associated with inversion symmetry breaking, which leads to non-vanishing hybridization of orbitals $d_{\pm}$ with the orbitals $d_{xy}$ and $d_{3z^2-1}$ on the nearest neighbor. In the tight-binding picture, Rashba Hamiltonian can be viewed as a second-order correction to the Bloch state energy associated with such nearest-neighbor interorbital hopping, followed by spin-flipping mediated by SOC~\cite{Petersen2000}.

Rashba contribution results in the Bychkov-Rashba effect~\cite{Rashba1984} $-$ splitting of the spin-degenerate band into two spin-momentum locked sub-bands, and Rashba-Edelstein effect~\cite{EDELSTEIN1990233} $-$ spin polarization of current driven by in-plane electric bias, which results from the difference between the densities of states in spin-momentum locked sub-bands. The latter can be interpreted as spin-polarization induced by the effective Rashba field $\mathbf{B}_R=\alpha_R\mu_B^{-1}\hat{z}\times\mathbf{\Delta k}$, where $\Delta k$ is the Fermi surface shift induced by electric bias.

In the proposed EDSR driven by Rashba SOC in STO with broken inversion symmetry, an in-plane ac electric field $\mathbf{E}\cos\omega t$ along the x-axis modulates the wavevector as $\mathbf{k}=-m_x\mu\mathbf{E}/\hbar$, where $\mu$ is the electron mobility, resulting in an effective ac spin-orbit field 
\begin{equation}\label{eq:B_eff}
	\mathbf{B}_R=-\frac{2\alpha_Rm^*}{\mu_B\hbar}\hat{\mathbf{z}}\times\mathbf{E}\cos\omega t.
\end{equation}
For $\mathbf{E}_0\parallel\mathbf{B}$, it acts as an ac magnetic field in the transverse ESR geometry, geometrically the same as shown in Fig.~\ref{fig:ESR_psi1}(a) but oscillating at the driving frequency. At cryogenic temperatures, lightly doped STO typically exhibits a high electron mobility of up to $\mu=10^4$~cm$^2$/V$\cdot$s~\cite{doi:10.1146/annurev-conmatphys-031218-013144}, enabling large modulations of wavevector by electric field. Rashba coefficients $\alpha_R$ of up to a few $meV\cdot\AA$ were reported for 2DEG at STO interface~\cite{omar2021large}. Using $\alpha_R=0.1$~meV$\cdot\AA$, we estimate that effective ac field amplitude $\approx10$~G comparable to that in the conventional ESR can be produced by a modest $E_0\approx1$~V/cm. At light doping, larger electric fields are likely achievable in micropatterned thin films without significant Joule heating.

Rashba-driven EDSR can be detected as a peak in imaginary component of impedance with respect to the driving ac voltage. AHE voltage detection discussed above for inversion-symmetric case can provide an effective alternative. In case of Rashba-driven EDSR, pseudospin precession occurs at the driving frequency, which according to Eq.~(\ref{eq:AHE}) results in mixing AHE current at $2\omega$ and a rectified dc current. Highly sensitive available measurement techniques of dc electrical signals allow efficient detection of the latter.

\section{Summary}\label{sec:summary}

In this work, we showed that cubic symmetry breaking in the tetragonal phase of strontium titanate (STO) and its epitaxially strained films results in a highly anomalous Zeeman effect of conduction-band electrons, due to the interplay between orbitally-selective anisotropy of the conduction sub-bands and spin-orbit coupling (SOC). Our main finding is that the spin-orbital composition of one of the two lowest-energy sub-bands is strongly dependent on lattice distortions, resulting in large variations of the g-factor. In particular, variation of strain-induced splitting between the two sub-bands by only $10$~meV can result in the variation of the g-factor between $0$ and almost $2$. The g-factor of this subband is expected to exhibit a large anisotropy even at small strain. In contrast, the second lowest-energy sub-band is expected to exhibit a very small g-factor insensitive to strain.

We also analyzed the possibility of all-electrical characterization of Zeeman effect by the electric dipole spin resonance (EDSR). This  approach is well-suited to STO due to the strong wavevector dependence of spin-orbital composition of the conduction band states, and the possibility of a large Rashba effect in heterostructures with broken inversion symmetry. In the latter case, it may be possible to detect the resonance as a rectified Hall voltage.

Our findings suggest that measurements of Zeeman effect in STO can resolve the ongoing debate about the ordering and spin-orbital composition of conduction sub-bands in STO, which may be central to the understanding of its unusual electronic properties. Measurements of the $g$-factor in the vicinity of the superconducting transition may help resolve the long-standing debate: is SC in STO phonon-mediated similarly to BCS~\cite{PhysRevResearch.4.013019}, or is it unconventional as in other complex oxides such as cuprates~\cite{PhysRevLett.121.127002}? Specifically, anomalous $g$-factors necessarily mean that the orbital composition of the Bloch states is spin- (and by extension momentum) dependent, and SOC has a substantial effect on the electronic states. Under these conditions, the isotropic single-band BCS approximation is not applicable, and the underlying mechanisms may be instead more closely related to the unconventional multi-orbital superconductors such as pnictides~\cite{Chen2008,Andersen2011} and ruthenates~\cite{PhysRevB.73.134501}. The latter possibility is supported by tunneling spectroscopy~\cite{PhysRevLett.45.1352} and thermal conductivity measurements~\cite{PhysRevB.90.140508}. This finding may have far-reaching implications for the classification of superconductors. BCS-like superconducting density of states such as that observed in STO~\cite{PhysRevB.92.174504} is commonly interpreted as a signature of the conventional mechanism. Instead, it may be a relatively generic consequence of electron pairing, which can be driven either by conventional or unconventional mechanisms such as repulsion~\cite{PhysRevB.106.224519}. Furthermore, observation of $g$-factor anisotropy close to the superconducting transition can provide insight into the mechanism of strong correlation between strain and superconductivity in STO~\cite{PhysRevB.97.144506,Ahadi2019}.

The predicted anomalous $g$-factor is a consequence of unquenched orbital moment permitted by the orbitally-selective band structure of STO. Consequently, this material may be uniquely suited for the generation and long-distance transport of orbital moment in orbitronic devices~\cite{Urazhdin2023} $-$ devices utilizing orbital moments to transmit and process information and to control the state of magnetic systems, which were proposed as an efficient alternative to analogous spin-based functionality in spintronics~\cite{Go2021-sn}.

This work was supported by the NSF award ECCS-2005786.

\bibliography{STOESR}
\bibliographystyle{apsrev4-2}
\end{document}